\documentclass[preprint,12pt]{elsarticle}

\usepackage{graphicx}
\usepackage{amssymb}

\usepackage{braket}
\usepackage{subfig}
\usepackage{adjustbox}
\usepackage{xcolor}
\usepackage{tabularx}
\usepackage{multirow}
\usepackage{amsmath}
\usepackage{float}
\usepackage{numcompress}

\usepackage{url}

\usepackage{breakurl}
\usepackage[breaklinks]{hyperref}

\usepackage{lineno}

\begin{document}

\begin{frontmatter}


\title{Quantum-Enhanced Machine Learning for Covid-19 and Anderson Insulator Predictions}



\author[McGill,Mila]{Paul-Aymeric McRae}
\author[McGill,INTRIQ]{Michael Hilke}

\address[McGill]{Department of Physics, McGill University, Montr\'eal, Canada H3A 2T8}
\address[Mila]{Mila - Quebec AI Institute, Montr\'eal, Qu\'ebec, Canada}
\address[INTRIQ]{INTRIQ - Quebec Institute for Quantum Information}

\begin{abstract}
Quantum Machine Learning (QML) algorithms to solve classifications problems have been made available thanks to recent advancements in quantum computation. While the number of qubits are still relatively small, they have been used for ``quantum enhancement" of machine learning. An important question is related to the efficacy of such protocols. We evaluate this efficacy using common baseline data sets, in addition to recent coronavirus spread data as well as the quantum metal-insulator transition in three dimensions. For the computation, we used the 16 qubit IBM quantum computer. We find that the ``quantum enhancement" is not generic and fails for more complex machine learning tasks.
\end{abstract}


\end{frontmatter}


\section{Introduction}
\label{S:1}

In the last decade or so, quantum computing has seen huge leaps in innovation and public awareness \cite{progress2019}. It has now become possible to run programs on real quantum computers \cite{ibm2019} or to realistically simulate them on a classical machine \cite{qasm}. Theoretically a quantum computer can offer great speed-up \cite{Ronnow2014} or better results \cite{Montanaro2016} as compared to algorithms performed on a classical binary system. Given these advances in the field it became interesting to combine it with another rapidly-expanding topic in computer science: machine learning, \cite{dunjko2018machine,Arunachalam2017,Ciliberto2018} leading to the aptly-named domain of Quantum Machine Learning (QML). Thanks to IBM's QISKit (Quantum Information Software Kit) \cite{ibm2019}, QML has become tangible and accessible \cite{havlicek}, with interesting results.

It is useful to define and classify different forms of QML. An effective representation is to consider the usual machine learning (ML) framework, where we have an input, I, a computation, C and an output O. In conventional ML all three (I, C and O) are performed classically. In QML, any of I, C, or O or any combination thereof can be quantum. Arguably, the simplest approach is to take a quantum system, for example an electron in a random potential (Anderson localization) \cite{Anderson}, which we label quI (quantum input) and use classical machine learning techniques to improve the classification boundary of localized and extended states \cite{ohtsuki2017deep}. Here we push the scheme one step further, by using a partial quantum C (or quC). In this classification scheme, both the input and the computation is partially quantum as we describe below. 

In contrast, it is possible to use a purely classical I, for example the Covid-19 infection rate in the US, but use a hybrid quantum C. In some cases, this approach was shown to lead to an improvement in the classification outcome, which was dubbed, ``quantum enhancement" \cite{dunjko2016quantum}. We use this scheme to explore the possible ``quantum enhancement" of different data samples, such as the more complex Covid-19 data mentioned above. In all cases, we consider a classical O, so that we only consider the cases I-quC-O and quI-quC-O as our QML framework. 

\subsection{Support Vector Machines}
The flavour of machine learning used here is a Support Vector Machine (SVM). SVMs are used to classify data into different categories (usually two) \cite{Cortes1995, scikit-learn, Rebentrost2014}. They are a type of supervised learning \cite{scikit-learn, Rebentrost2014}; that is, the branch of machine learning in which a already-classified data set is used as an input on which the machine trains \cite{Rebentrost2014}. Upon successful training, the SVM can then predict the category in which a new data point belongs \cite{Rebentrost2014,mehryarmohri2012,stuartrussell2009}. 

Support Vector Machines work by taking the training data set and virtually plotting the data in $d$-dimensional space where $d$ is the number of parameters (or features) of each data point. It then attempts to find the $(d-1)$-dimensional hyperplane which separates the data into two classes (in the case of data with two parameters this is a line) \cite{berwick2003idiot}. Specifically, it aims to find the hyperplane with the largest margin between points of either class \cite{scikit-learn, berwick2003idiot}. Points lying on this margin are known as support vectors \cite{berwick2003idiot}.

In many cases it is impossible to draw a separating hyperplane between classes of data, especially when the boundary is highly nonlinear \cite{berwick2003idiot}. In such a case where the SVM is unable to find a suitable separating hyperplane, a technique known as a ``kernel trick'' is used, which involves using a non-linear function (often called a Feature Map \cite{Havlek2019}) to project the data into a higher dimension where a separating hyperplane may be found \cite{Boser1992}.

In this work, all classical SVM operations have been performed using the SVM Python module developed by scikit-learn, whose documentation can be found in \cite{scikit-learn}. 

\begin{figure}
\centering
\includegraphics[width=0.8\linewidth]{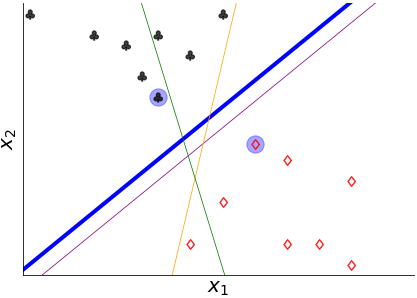}
\caption{Two-parameter ($x_1$, $x_2$) data set with two categories (clubs and diamonds) along with possible separating hyperplanes. The green line does not separate the  data into two sets. The orange and purple lines do separate the data but do not do so maximally. The blue line is the maximally separating hyperplane with support vectors in blue circles.}
\label{fig:svm}
\end{figure}

\subsection{Quantum Support Vector Machines}
The quantum version of the SVM is best described as ``quantum-assisted" or ``quantum-enhanced" \cite{Havlek2019} in the sense that the algorithm is largely classical with certain operations performed by a quantum processor (real or simulated). The Quantum SVM (or QSVM) made available by IBM QISKit functions the same way as a classical SVM but the feature mapping \cite{Havlek2019} and some operations of the hyperplane calculation \cite{Rebentrost2014} are handled by the quantum processor.

\subsection{Variational Quantum Classifier}
An important quantum alternative to the QSVM is the Variational Quantum Classifier (VQC). This belongs to a class known as Variational Quantum Algorithms, which are hybrid classical-quantum algorithms which iteratively use the measurement of a parametrized quantum circuit as the input of a classical process. The classical algorithm's output is then passed as the parameter of the quantum circuit, and the loop then repeats \cite{LaRose2019, DhruvSaldanha2020}. The VQC encodes a data input as a quantum state, which is then processed and the quantum probability is then measured, which serves as the base inference for the binary classification. By refining this inference, the VQC can make predictions based on data.

\begin{figure}[H]
   \centering
   \subfloat[][]{\includegraphics[width=.7\textwidth]{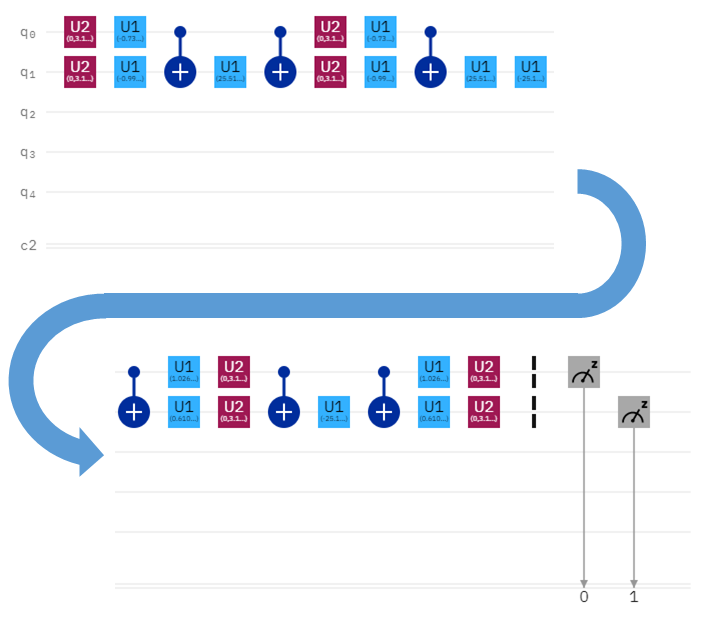}}\\
   \subfloat[][]{\includegraphics[width=.9\textwidth]{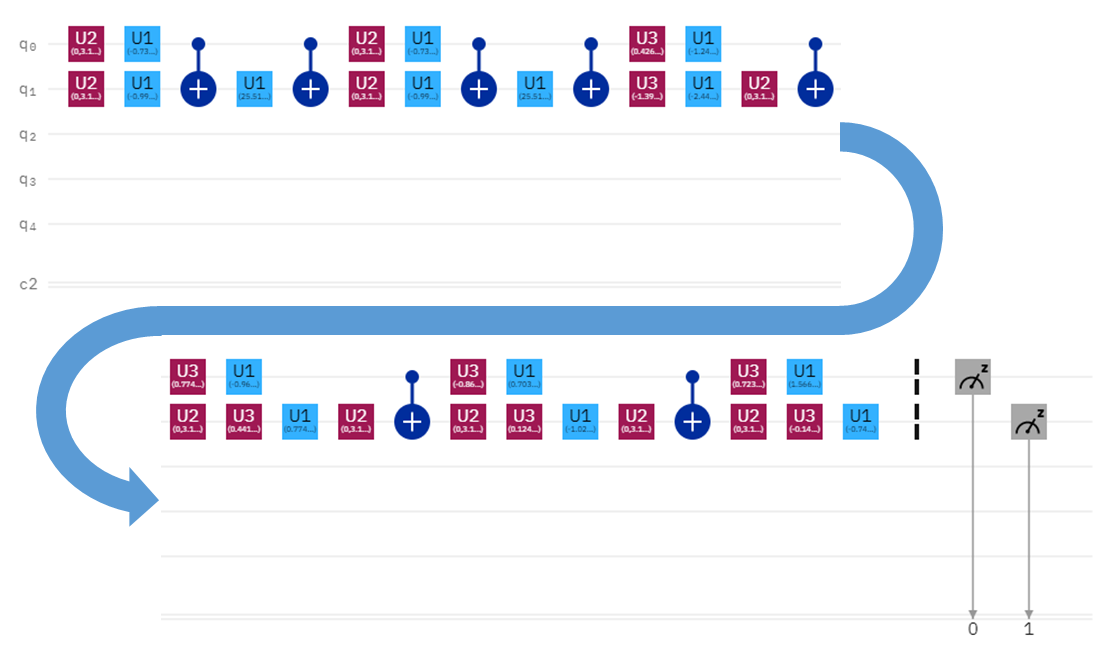}}\\
   \caption{Sample circuits from (a) QSVM and (b) VQC runs on a 5-qubit system. Wrap-around for visualization.}
   \label{fig:Circuits}
\end{figure}

\subsection{IBM Quantum Experience and QISKit}
The term ``quantum computing" can refer to several different types of machines \cite{progress2019}. Here we refer specifically to those publicly available via the IBM Quantum Experience. These operate on the model of quantum circuits \cite{QiskitCircuit}, a system which can be defined as follows: a classical computer stores information in bits which can hold a value of either 0 or 1, and can perform operations on these bits using a combination of logic gates (specifically AND, OR, NOT) \cite{davidpatterson2013}. By analogy a quantum circuit uses qubits which can store superpositions of $\ket{0}$ and $\ket{1}$ states and operates on them using quantum gates \cite{Qiskit-Textbook}, which in quantum mechanical terms can be described by operators on two-level quantum systems and product states of such systems \cite{nielsen2001quantum}.

IBM has developed cloud-based access to real quantum circuit-based computers as part of the IBM Quantum Experience \cite{ibmnewsroom2016} for which software can be written using QISKit \cite{progress2019}. Several backends are available, with 5 \cite{5qubit} or 16 qubits \cite{16qubit} or a simulator of over 30 qubits \cite{ibmsimulator}. Their real computers are built using transmons \cite{Gambetta2017}, a type of superconducting charge qubit \cite{Kjaergaard_2020,Koch_2007}, with algorithms designed to reduce errors that arise due to the fragility of quantum information \cite{ibmerrors,Chow2014}.

\subsection{Sources of Data}
For our work, we use publicly available data sets provided by UC Irvine as well as computed data. The first UCI data set is the ``Breast Cancer Wisconsin (Diagnostic) Data Set", which contains 569 instances of tumor data each containing 32 attributes which are used to determine whether the tumor is benign or malignant. The second UCI set is known as ``Wine Data Set" and contains 178 instances of data sourced from wine, each containing 13 attributes which are used to classify which of three cultivars the wine was sourced from \cite{Dua:2019}. These two data sets are often used as baselines for QML trials, including in \cite{havenstein_thomas_chandrasekaran_2018} as well as in IBM's own QISKit tutorials \cite{Qiskit}.

Our generated data falls into three classes: ad-hoc data, Anderson data and Coronavirus data. The ad-hoc data describes all data generated on a 20 by 20 grid, some of which use Python's built-in pseudorandom functions. The coronavirus data used was compiled by The New York Times and publicly released on GitHub \cite{COVID}. It contains the cumulative number of cases in the United states on a county and state level. For this, the county-level data is used. Finally the Anderson data is based on Anderson's seminal result, where he showed that certain materials can undergo a quantum phase transition between conductor and an insulator based on the strength of disorder in the system \cite{Anderson}. This can be numerically simulated \cite{Croy} by varying energy and disorder to get a two-dimensional data picture. The data is generated using an iterative Green's function technique to obtain the transmission coefficient as a function of disorder strength \cite{datta2000nanoscale}.

\section{Methods}

\subsection{Classification}

As SVMs are designed to classify data into one of two sets (in most cases) based on a training input, all data used needed some parameter that delineates the set to which a data point would belong. The training input is defined by taking the entirety of the classified data set and using 67\% to fill the training set. The remaining 33\% is used as a ``testing set" to verify the accuracy of the SVM. The classifier used per data type is described as follows.

For both the ``Wine" and ``Breast Cancer" data sets, QISKit provides functions to procure the data such that it is already classified, and each point can be reduced in dimensionality using the PCA algorithm. In our experiments, we primarily reduced the feature dimension to two, though certain experiments use three- or four-dimensional feature data.

\begin{figure}[H]
   \centering
   \subfloat[][]{\includegraphics[width=.4\textwidth]{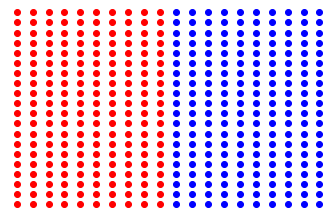}}\quad
   \subfloat[][]{\includegraphics[width=.4\textwidth]{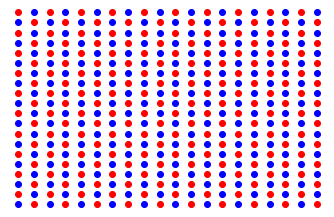}}\\
   \subfloat[][]{\includegraphics[width=.4\textwidth]{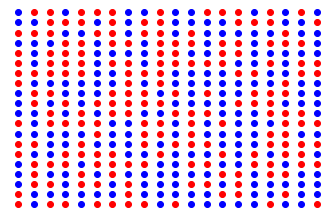}}\quad
   \subfloat[][]{\includegraphics[width=.4\textwidth]{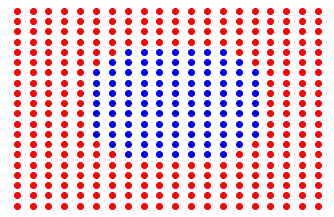}}
   \caption{Ad-Hoc data. (a) Perfectly linearly separable data; (b) Alternating category between adjacent points; (c) Randomly assigned points; (d) ``Circle" of one category surrounded by the other}
   \label{fig:adhoc}
\end{figure}

In the case of the ad-hoc data the classifier is defined along with the data, with the goal to create a particular ``shape". As mentioned, all data was generated on a $20 \times 20$ grid with increment size 1. The four different generated data sets are presented in Figure \ref{fig:adhoc}.

The coronavirus example uses data points consisting of a US county's population and its total number of cases, per day since the first recorded case in the country. The classifier was whether a county is deemed ``safe" or ``unsafe" based on whether the famous ``curve" (i.e. the number of new cases per day) was sufficiently flat or in fact decreasing. This was done by observing the curve locally by taking the change in the number of cases over three days preceding May 8th (an arbitrary choice). The derivative of cases per day over the three days was then averaged. If this average was below a certain threshold, i.e. the curve was flat or decreasing, the county was deemed safe. Otherwise, it is classified as unsafe. The ML experiments were run using thresholds of 1, 3, and 5 new cases per day on average. Only counties which had a nonzero number of cases over the May $5^{th}$ to $8^{th}$ interval were included in the data set. Likewise counties which changed the geographical area of reported data over this time period were excluded. 

\begin{figure}[H]
   \centering
   {\includegraphics[width=\textwidth]{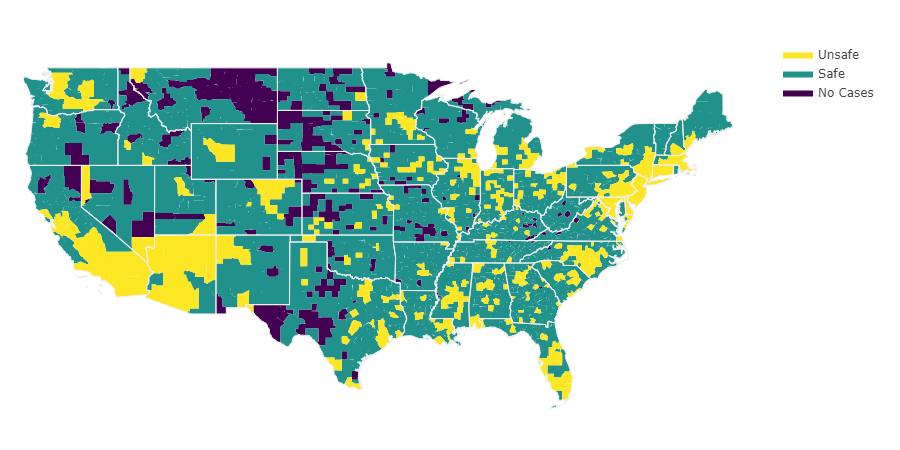}}
   \caption{Coronavirus data. Yellow counties are unsafe (curve not flattened), teal counties are safe (curve flattened) and purple countries have no cases or insufficient data. We define ``Safe" as $\leq 5$ new cases per day on average.}
   \label{fig:covid}
\end{figure}

The Anderson data was generated numerically by considering an $L\times L\times L$ cube between two semi-infinite leads. The conductivity of the material is computed by the transmission coefficient via the Landauer formula as a function of Energy ($E$), disorder ($V_a$), cube side length ($L$) for different disorder configurations. The transmission coefficient is obtained from the Green's function by tracing over all channels \cite{datta2000nanoscale}. The full Green's function is computed using a layer by layer iterative computation of the partial Green's function \cite{mackinnon1985calculation}. The conductance (which we denote $G$) was calculated for varying values of the parameters ($E$, $L$, $V_a$ and disorder configurations). We used 100 different disorder configurations for the averaging and a maximum of $L=14$. The dimensionality of this matrix was reduced by log-averaging the value of $G$ for all disorder configurations for fixed $E$, $V_a$, and $L$, leading to $\bar{G}$. The following finite-size scaling method was used to differentiate metallic and insulating states \cite{mackinnon1981one}. A least-squares linear fit to approximate $\bar{G}$ as a function of $L$, whose slope (or partial derivative $\frac{\partial \bar{G}}{\partial L}$) was recorded for each $E$ and $V_a$. The classification was then made based on this value: a positive $\frac{\partial \bar{G}}{\partial L}$ means the system is conductive, while a negative number indicates the system is an insulator. Values within $1 \times 10 ^{-6}$ of $0$ were ignored.

\begin{figure}[H]
   \centering
   \subfloat[][]{\includegraphics[width=.5\textwidth]{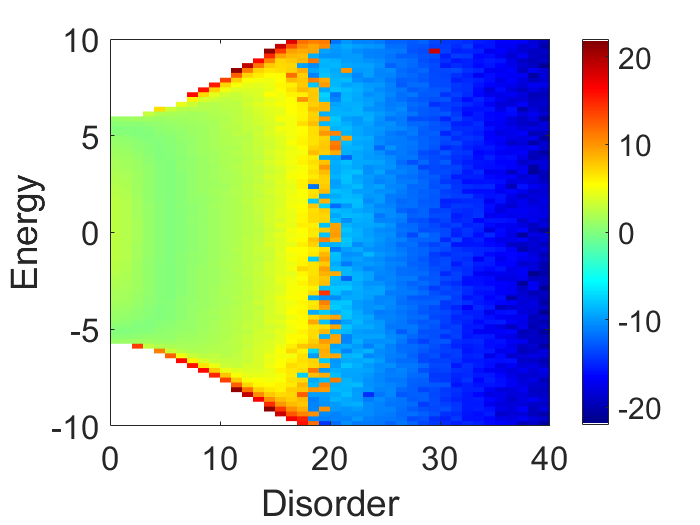}}\quad
   \subfloat[][]{\includegraphics[width=.42\textwidth]{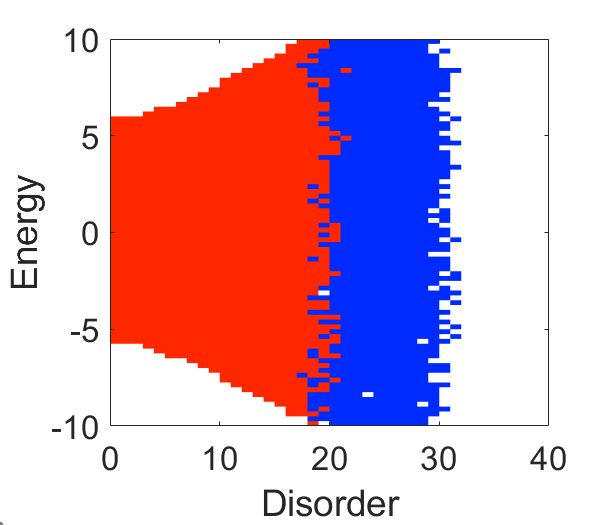}}\\
   \caption{Anderson data: Plot (a) shows the raw generated data where the color scale corresponds to $\mbox{sgn}(\partial_LG)|\log|\partial_LG||$ and plot (b) is the classified data: $\mbox{sgn}(\partial_LG)$ (red corresponds to a conductor, blue to an insulator, and white was excluded with values $|\partial_LG)|<10^{-6}$ ).}
   \label{fig:Anderson}
\end{figure}

\subsection{Randomization}

To achieve an apparently random state, two principles were kept in mind: reproducibility and invertibility. As such, ``randomization" was achieved using custom made Pseudo-Random Number Generators (PRNG), which generate numbers that appear random but whose behaviour can actually be predicted \cite{TCS010}. Specifically given the same input and the same defining parameter (dubbed the ``seed") a PRNG will always return the same value \cite{TCS010}. Here, each of the two parameters for each data point was run through the same PNRG, though different problems used different generators.

We can define a PRNG as follows:

\begin{equation}
    rand(x) = 1+ (a \cdot x + b)\mod m
\end{equation}

Where $b$ is an arbitrary integer, and $a$ and $m$ are prime numbers, with $a < m$ and $m$ chosen such that it is greater than all the values of the data, in order to guarantee invertibility. The inverse can be found as:

\begin{equation}
    rand^{-1}(y) = a^{-1} \cdot (y - 1 - b) \mod m
\end{equation}

Where $a^{-1}$ is the modular inverse of $a$ (that is, the value such that $a\cdot a^{-1} \mod m \equiv 1$. This value can be found by iterating through all integers between $1$ and $m$ until a number satisfying this condition is found. If $a$ and $m$ are coprime (which they must be since they are both prime) then this inverse must exist and is unique \cite{shoup2005a}.

\subsection{Data Reduction}

During the experimentation, we developed a technique to reduce the size of the data set. Notably this had the benefit of removing outliers, which we observed increased the accuracy. As well, with a smaller data set the training time would be reduced. To achieve this reduction it was important to keep data point more relevant to the training (that is, points closer to the data boundary where classification becomes more difficult). This was achieved by using a classical linear SVM to draw the hyperplane that best separates the data into two classes, then selecting data points that fall within some distance of this hyperplane. The chosen distance was tuned manually for best results.

\subsection{Machine Learning Procedures}

Classical machine learning experiments were done using the scikit-learn Python package \cite{scikit-learn} and specifically the SVM module. Most experiments were done ``out of the box" by using a linear kernel, but occasionally the machine was tweaked by tuning the hyperparameters, specifically by changing the kernel to a polynomial or RBF. In some trials, the data was preprocessed using a standard scaler, which rescales the data by subtracting the mean of the data from the sample and dividing the result by the standard deviation. This drastically decreases the training time, especially on large data sets. As a rule, on data which was randomized, unscaled data would not be tested as the algorithm would not converge in reasonable time.

All quantum machine learning was done using IBM QISKit's QSVM module \cite{QSVM}. The feature map used was Second-order Pauli-Z evolution circuit \cite{ZZFeatureMap} using two repeated circuits and linear entanglement. QML experiments were performed on two backends: the QASM simulator \cite{qasm} and the 16-qubit computer in Melbourne \cite{16qubit}.

\section{Results and Analysis}

\subsection{Volatility of Results}
\label{subsection:volatility}

Havenstein et. al. \cite{havenstein_thomas_chandrasekaran_2018} show that a VQC can outperform a classical SVM on both the Wine and Breast Cancer data sets, a result we were unable to reproduce. This can be accounted by the volatility of running similar, but not identical trials. For instance we notice that accuracy differences can arise from minor changes in the way the data is partitioned into training/test data, including the seed (called state in the code) used to pseudo-randomly shuffle the data as well as the size of the sets. We report these difference in Table \ref{table:volatility}. 

\begin{table}[H]
\caption{Accuracy of different runs using slightly different train/test partitions for the QASM simulator (first table) and the IBMQ 16-Qubit Melbourne Backend (second table).}
    \begin{tabular}{l|l}
    \textbf{Data Partitioning}                             & \textbf{Accuracy (\%) }\\ \hline
    Using 33-37 test/train split, random state 42 & 75.0     \\
    Using 30-70 test/train split, random state 42 & 72.22    \\
    Using 33-67 test/train split, random state 12 & 75.0     \\
    Using 30-70 test/train split, random state 12 & 66.67   
    \end{tabular}
    \newline
    \vspace*{0.5 cm}
    \newline
    \begin{tabular}{l|l}
    \textbf{Data Partitioning                             }& \textbf{Accuracy (\%) }\\ \hline
    Using 33-37 test/train split, random state 42 & 75.0     \\
    Using 30-70 test/train split, random state 42 & 85.0   \\
    Using 33-67 test/train split, random state 12 & 66.67     \\
    Using 30-70 test/train split, random state 12 & 72.22   
    \end{tabular}
\label{table:volatility}
\end{table}

We also note that on successive trials run on either a real quantum backend or the statevector simuluator, accuracy would fluctuate, which makes the outputs of these machines stochastic and non-reproducible.

\subsection{Time Complexity and Runtime}

\begin{figure}
\centering
\includegraphics[width=0.8\linewidth]{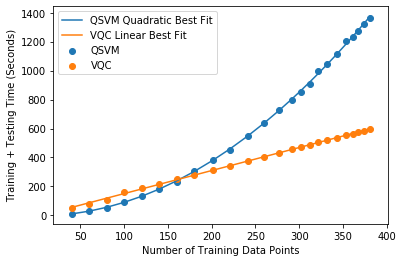}
\caption{Computation time for quantum classification algorithms versus number of training data points, along with least-squares fits.}
\label{fig:bc_time}
\end{figure}

\begin{figure}
\centering
\includegraphics[width=0.9\linewidth]{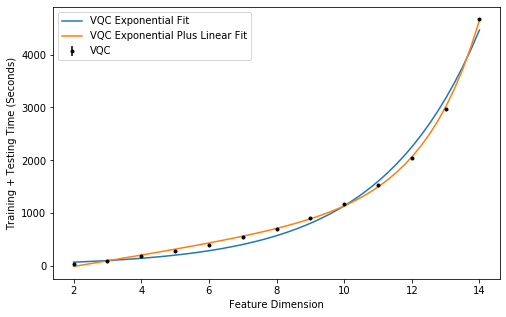}
\caption{Computation time for VQC versus feature dimension. We have both the exponential evolution (which is captured by big-O, which drops slowly increasing terms) and the closer fit: an exponential plus a linear function.}
\label{fig:VQC_time}
\end{figure}

While experimenting with QML we noted very long training and prediction times when using large data sets. Naturally this raised the question of how runtimes scale with the size of the input. We observed that the duration of a QSVM run on the QASM Simulator scaled as $O(n^2)$ where $n$ is the size of the training data (Fig. \ref{fig:bc_time}). We similarly observed linear $O(n)$ time using a VQC with a QASM simulator (Fig. \ref{fig:bc_time}) as well as an exponential $O(2^d)$ time dependence on the feature dimension $d$ (see Figure \ref{fig:VQC_time}). The simulator was used rather than real backend since the time to run experiments on them depends on the number of jobs in the queue which is a factor beyond our control. We demonstrate the time dependence in the figures using least-squares fits.

As presented in an overview by \cite{DUAN2020126595}, the theorized time complexity of the QSVM algorithm is $\tilde O(log(nd)\kappa^2/\epsilon^3)$ where $\kappa$ is a quantity dependent on the kernel (specifically, it is the condition number of a matrix which must be inverted), $\epsilon$ is the accuracy, and we use the notation $\tilde O(h(n))= O(h(n)log(h(n))$ (i.e. suppressing slowly varying multiplicative terms). This notably differs from our observation, which is likely to arise from two factors: $(1)$ the fact that a simulator was used and not a pure quantum circuit and $(2)$ the difference between implementation of an algorithm and theoretical performance. Expanding on this second point, consider the analogous example of binary search, which can be completed in $O(log(n))$ time, but only if the list is sorted. If not, the list must be sorted first, and this overhead will cost at least $O(n log(n))$ time.

By exploring the implementation of the QSVM we can see the $O(n^2)$ comes from computation of the kernel matrix; that is, the matrix of the kernel function computed between all pairs of data points. Since for $n$ datapoints there are $n^2$ such pairs, the origin of the quadratic computational complexity is obvious. The VQC's linear dependence on $n$ is harder to explain, though a heuristic explanation is the absence of any quadratic-time functions and the fact that the algorithm likely makes at least one pass over all the data. We were, however, unable to explain the exponential dependence on feature dimension.

\subsection{Baselines Revisited}

A quick search through literature or online documentation will reveal that the Wine and Breast Cancer data sets are commonly used to test QML algorithms \cite{sharma2020qeml, havenstein_thomas_chandrasekaran_2018, mendozza_acampora_vitiello_2019}. However, these are often run without comparisons to a classical equivalent, such as the case of the QISKit tutorials \cite{Qiskit} or the thesis by Mendozza \cite{mendozza_acampora_vitiello_2019}. Others, such as Havenstein et. al. \cite{havenstein_thomas_chandrasekaran_2018} make explicit comparisons, though we were not able to reproduce their exact results for reasons stated in Subsection \ref{subsection:volatility}.

Here we present our findings for our experimentation, with varying parameters such as feature dimension, number of data points, and backend used. In no experiments using either the Breast Cancer or Wine data sets were we able to observe a better accuracy of a quantum method as compared to classical counterparts. This would seem to indicate that claims of a quantum advantage using these baselines may be overstated, and that a new baseline should be explored which more clearly presents a quantum advantage.

\begin{figure}
\centering
\includegraphics[width=0.8\linewidth]{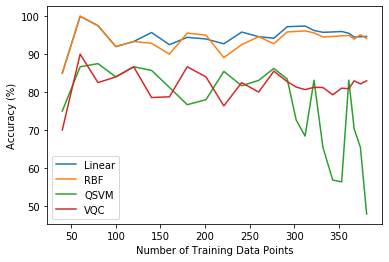}
\caption{Accuracy of various classification methods versus number of training samples from the Breast Cancer data set. The classical methods (Linear and RBF kernel SVMs) consistently outperform the QML.}
\label{fig:bc_accuracy}
\end{figure}

\subsection{Results on Coronavirus and Anderson}

While using a benchmark data set to reveal details on the performance of a machine learning algorithm, the most important question when exploring a novel technique is how well does it work in practice. For this we used two data sets from ``real" sources (Coronavirus and Anderson data) as well as several manufactured data distributions (which we refer to as Ad-Hoc). In this subsection we present our conclusions from this data. Our raw results can be found in \ref{S:AppendixB}.

Here again we note that in nearly every case, a form of classical classification outperforms a quantum counterpart. In some rare instances, quantum narrowly displays better accuracy than classical algorithms, generally when the setup has a degree of randomness involved. On these data sets we observe the larger prevalence of a potentially catastrophic behaviour: some of the quantum methods achieve an artificially high accuracy by solely predicting one class. This generally occurs when the data disparity is skewed towards one class. However, unlike the classical methods implemented in scikit-learn, QISKit classification algorithms do not allow the same degree of hyperparameter tuning, which makes this behaviour potentially unavoidable.

\section{Conclusion}

While quantum computation is a field of rapidly evolving interest, the practicality of quantum classification algorithms lags behind that of simple classical methods. Lower accuracy and longer times, caused by the high complexity of simulation and the relatively limited real backends, have caused QML to be a novelty more than a great leap in innovation. Nevertheless, with refinement of existing algorithms, including more variability in the tuning of hyperparameters, QML may in the future emerge as a strong contender in the field of machine learning.



\bibliography{sample}

\appendix

\section{RBF  kernel  classical  SVM}

An RBF kernel classical SVM without data preprocessing can be used to locate the data boundary between the insulator/conductor phase transition in the Anderson problem to 96.67\% accuracy. Even when the data was reduced and the boundary became less visibly obvious, this method of classification was able to obtain 79.7\% accuracy. We can conclude that this is a good metric for classifying the properties of a a disordered solid.

We also observed that classically, using an RBF kernel, we were able to predict whether a county in the USA was deemed safe to 90.99\% accuracy. A linear kernel yielded a slightly smaller accuracy at 90.56\%. As a verification of the quality of this process, we used all data collected from the 8th of May as a training set and tested the machine on several other dates, with both RBF and linear kernels, as reported in Table \ref{table:covidpred}. We observe that although RBF is incrementally better on present-day data, the predictive capacity of the linear kernel surpasses it. We also note the linear kernel's decaying prediction accuracy, which we can likely attribute to the data drifting further from its initial state.

\begin{table}[H]
\centering
\caption{Prediction accuracy of a machine trained using May 8th COVID data.}
\begin{tabular}{lllll}
                            & May 9   & May 18  & May 28  & June 8  \\ \cline{2-5} 
\multicolumn{1}{l|}{Linear} & 91.55\% & 91.07\% & 90.20\% & 88.34\% \\
\multicolumn{1}{l|}{RBF}    & 81.01\% & 82.61\% & 81.84\% & 82.72\%
\end{tabular}
\label{table:covidpred}
\end{table}

\section{Data Tables}
\label{S:AppendixB}

\begin{table}[H]
\tiny
\centering
\begin{tabularx}{\textwidth}{l|l|X|X|X}
\textbf{Algorithm} & \textbf{Backend} & \textbf{Time (Seconds, rounded)} & \textbf{Accuracy} & \textbf{Only one Class Predicted?} \\ \hline
SVM, Linear Kernel & Local Processor & 0 & 85.0 & False \\
SVM, Linear Kernel, scaled & Local Processor & 0 & 85.0 & False \\
SVM, RBF Kernel & Local Processor & 0 & 85.0 & False \\
SVM, RBF Kernel, scaled & Local Processor & 0 & 85.0 & False \\
QSVM & qasm\_simulator & 11 & 75.0 & False \\
VQC & qasm\_simulator & 48 & 70.0 & False \\
QSVM & ibmq\_16\_melbourne & 1966 & 75.0 & False \\
QSVM & statevector\_simulator & 0 & 75.0 & False \\
VQC & statevector\_simulator & 31 & 75.0 & False
\end{tabularx}
\caption{Results of trials on breast cancer data set with 60 total samples. Trials run on ibmq\_16\_melbourne and statevector\_simulator are the results from one trial and were not necessarily reproducible.}
\label{table:Breast_cancer}
\end{table}

\begin{table}[H]
\tiny
\begin{tabularx}{\textwidth}{c|l|l|X|X|X}
\textbf{Feature Dimension} & \textbf{Algorithm} & \textbf{Backend} & \textbf{Time (Seconds, rounded)} & \textbf{Accuracy (\%)} & \textbf{Only one Class Predicted?} \\ \hline
\multirow{4}{*}{2} & SVM, Linear Kernel, scaled & Local Processor & 0 & 100.0 & False \\
 & SVM, RBF Kernel, scaled & Local Processor & 0 & 100.0 & False \\
 & QSVM & qasm\_simulator & 6 & 100.0 & False \\
 & VQC & qasm\_simulator & 32 & 71.43 & False \\ \hline
\multirow{4}{*}{3} & SVM, Linear Kernel, scaled & Local Processor & 0 & 100.0 & False \\
 & SVM, RBF Kernel, scaled & Local Processor & 0 & 92.86 & False \\
 & QSVM & qasm\_simulator & 14 & 64.29 & False \\
 & VQC & qasm\_simulator & 58 & 42.86 & False \\ \hline
\multirow{6}{*}{4} & SVM, Linear Kernel & Local Processor & 0 & 92.86 & False \\
 & SVM, Linear Kernel, scaled & Local Processor & 0 & 92.86 & False \\
 & SVM, RBF Kernel & Local Processor & 0 & 100.0 & False \\
 & SVM, RBF Kernel, scaled & Local Processor & 0 & 100.0 & False \\
 & VQC & qasm\_simulator & 128 & 50.0 & False \\
\end{tabularx}
\caption{Results of trials of the Wine data set, with increasing feature dimension. Feature dimensions higher than 3 are not supported for QSVMs.}
\end{table}

\begin{table*}[!htbp]
\tiny
\begin{tabularx}{\textwidth}{c|l|l|X|X|X}
\multicolumn{1}{c}{\textbf{Data}} & \textbf{Algorithm} & \textbf{Backend} & \textbf{Time (Seconds, rounded)} & \textbf{Accuracy (\%)} & \textbf{Only one Class Predicted?} \\ \hline
\multirow{6}{*}{All Data} & SVM, Linear Kernel & Local Processor & 0 & 100.0 & False \\
 & SVM, Linear Kernel, scaled & Local Processor & 0 & 100.0 & False \\
 & SVM, RBF Kernel & Local Processor & 0 & 100.0 & False \\
 & SVM, RBF Kernel, scaled & Local Processor & 0 & 100.0 & False \\
 & QSVM & qasm\_simulator & 445 & 43.18 & False \\
 & VQC & qasm\_simulator & 315 & 54.55 & False \\ \hline
\multirow{6}{*}{\begin{tabular}[c]{@{}c@{}}All Data, \\ Randomized\end{tabular}} & SVM, Linear Kernel & Local Processor & 16 & 50.76 & False \\
 & SVM, Linear Kernel, scaled & Local Processor & 0 & 59.85 & False \\
 & SVM, RBF Kernel & Local Processor & 0 & 50.0 & True \\
 & SVM, RBF Kernel, scaled & Local Processor & 0 & 46.21 & False \\
 & QSVM & qasm\_simulator & 444 & 50.0 & False \\
 & VQC & qasm\_simulator & 314 & 54.55 & False
\end{tabularx}
\caption{Results for trials using perfectly linearly separable Ad-Hoc data.}
\end{table*}

\begin{table*}[!htbp]
\tiny
\begin{tabularx}{\textwidth}{c|l|l|X|X|X}
\multicolumn{1}{c}{\textbf{Data}} & \textbf{Algorithm} & \textbf{Backend} & \textbf{Time (Seconds, rounded)} & \textbf{Accuracy (\%)} & \textbf{Only one Class Predicted?} \\ \hline
\multirow{6}{*}{All Data} & SVM, Linear Kernel & Local Processor & 0 & 50.0 & False \\
 & SVM, Linear Kernel, scaled & Local Processor & 0 & 50.0 & False \\
 & SVM, RBF Kernel & Local Processor & 0 & 20.45 & False \\
 & SVM, RBF Kernel, scaled & Local Processor & 0 & 53.03 & False \\
 & QSVM & qasm\_simulator & 433 & 48.48 & False \\
 & VQC & qasm\_simulator & 301 & 52.27 & False \\ \hline
\multirow{6}{*}{\begin{tabular}[c]{@{}c@{}}All Data, \\ Randomized\end{tabular}} & SVM, Linear Kernel & Local Processor & 17 & 47.73 & False \\
 & SVM, Linear Kernel, scaled & Local Processor & 0 & 45.45 & False \\
 & SVM, RBF Kernel & Local Processor & 0 & 50.0 & True \\
 & SVM, RBF Kernel, scaled & Local Processor & 0 & 42.42 & False \\
 & QSVM & qasm\_simulator & 423 & 48.48 & False \\
 & VQC & qasm\_simulator & 300 & 62.12 & False
\end{tabularx}
\caption{Results for trials using alternating grid Ad-Hoc data.}
\end{table*}

\begin{table*}[!htbp]
\tiny
\begin{tabularx}{\textwidth}{c|l|l|X|X|X}
\multicolumn{1}{c}{\textbf{Data}} & \textbf{Algorithm} & \textbf{Backend} & \textbf{Time (Seconds, rounded)} & \textbf{Accuracy (\%)} & \textbf{Only one Class Predicted?} \\ \hline
\multirow{6}{*}{All Data} & SVM, Linear Kernel & Local Processor & 0 & 52.63 & True \\
 & SVM, Linear Kernel, scaled & Local Processor & 0 & 52.63 & True \\
 & SVM, RBF Kernel & Local Processor & 0 & 52.63 & False \\
 & SVM, RBF Kernel, scaled & Local Processor & 0 & 46.62 & False \\
 & QSVM & qasm\_simulator & 424 & 49.62 & False \\
 & VQC & qasm\_simulator & 299 & 51.13 & False \\ \hline
\multirow{6}{*}{\begin{tabular}[c]{@{}c@{}}All Data, \\ Randomized\end{tabular}} & SVM, Linear Kernel & Local Processor & 33 & 54.14 & False \\
 & SVM, Linear Kernel, scaled & Local Processor & 0 & 52.63 & True \\
 & SVM, RBF Kernel & Local Processor & 0 & 52.63 & True \\
 & SVM, RBF Kernel, scaled & Local Processor & 0 & 50.38 & False \\
 & QSVM & qasm\_simulator & 422 & 50.38 & False \\
 & VQC & qasm\_simulator & 299 & 49.62 & False
\end{tabularx}
\caption{Results for trials using random grid Ad-Hoc data.}
\end{table*}

\begin{table*}[!htbp]
\tiny
\begin{tabularx}{\textwidth}{c|l|l|X|X|X}
\multicolumn{1}{c}{\textbf{Data}} & \textbf{Algorithm} & \textbf{Backend} & \textbf{Time (Seconds, rounded)} & \textbf{Accuracy (\%)} & \textbf{Only one Class Predicted?} \\ \hline
\multirow{6}{*}{All Data} & SVM, Linear Kernel & Local Processor & 0 & 72.93 & True \\
 & SVM, Linear Kernel, scaled & Local Processor & 0 & 72.93 & True \\
 & SVM, RBF Kernel & Local Processor & 0 & 97.74 & False \\
 & SVM, RBF Kernel, scaled & Local Processor & 0 & 90.98 & False \\
 & QSVM & qasm\_simulator & 423 & 72.93 & True \\
 & VQC & qasm\_simulator & 299 & 57.89 & False \\
\multirow{6}{*}{\begin{tabular}[c]{@{}c@{}}All Data, \\ Randomized\end{tabular}} & SVM, Linear Kernel & Local Processor & 42 & 66.17 & False \\ \hline
 & SVM, Linear Kernel, scaled & Local Processor & 0 & 72.93 & True \\
 & SVM, RBF Kernel & Local Processor & 0 & 72.93 & True \\
 & SVM, RBF Kernel, scaled & Local Processor & 0 & 72.93 & True \\
 & QSVM & qasm\_simulator & 422 & 72.93 & True \\
 & VQC & qasm\_simulator & 300 & 54.89 & False
\end{tabularx}
\caption{Results for trials using ``circle" Ad-Hoc data.}
\end{table*}

\begin{table*}[!htbp]
\tiny
\begin{tabularx}{\textwidth}{c|l|l|X|X|X}
\multicolumn{1}{c}{\textbf{Data}} & \textbf{Algorithm} & \textbf{Backend} & \textbf{Time (Seconds, rounded)} & \textbf{Accuracy (\%)} & \textbf{Only one Class Predicted?} \\ \hline
\multirow{6}{*}{All Data} & SVM, Linear Kernel & Local Processor & 0 & 95.07 & False \\
 & SVM, Linear Kernel, scaled & Local Processor & 0 & 95.51 & False \\
 & SVM, RBF Kernel & Local Processor & 0 & 96.67 & False \\
 & SVM, RBF Kernel, scaled & Local Processor & 0 & 96.0 & False \\
 & QSVM & qasm\_simulator & 14009 & 56.52 & False \\
 & VQC & qasm\_simulator & 1722 & 49.42 & False \\ \hline
\multirow{4}{*}{\begin{tabular}[c]{@{}c@{}}All Data, \\ Randomized\end{tabular}} & SVM, Linear Kernel, scaled & Local Processor & 0 & 49.28 & False \\
 & SVM, RBF Kernel, scaled & Local Processor & 0 & 93.0 & False \\
 & QSVM & qasm\_simulator & 14042 & 54.64 & False \\
 & VQC & qasm\_simulator & 1707 & 54.2 & False \\ \hline
\multirow{6}{*}{Reduced Data} & SVM, Linear Kernel & Local Processor & 0 & 76.69 & False \\
 & SVM, Linear Kernel, scaled & Local Processor & 0 & 78.2 & False \\
 & SVM, RBF Kernel & Local Processor & 0 & 79.7 & False \\
 & SVM, RBF Kernel, scaled & Local Processor & 0 & 77.0 & False \\
 & QSVM & qasm\_simulator & 493 & 54.14 & False \\
 & VQC & qasm\_simulator & 322 & 60.9 & False \\ \hline
\multirow{6}{*}{\begin{tabular}[c]{@{}c@{}}Reduced Data, \\ Randomized\end{tabular}} & SVM, Linear Kernel, scaled & Local Processor & 0 & 50.38 & False \\
 & SVM, RBF Kernel, scaled & Local Processor & 0 & 60.0 & False \\
 & QSVM & qasm\_simulator & 485 & 60.15 & False \\
 & VQC & qasm\_simulator & 323 & 59.4 & False \\
 & QSVM & ibmq\_16\_melbourne & 117525 & 57.14 & False \\
 & QSVM & ibmq\_16\_melbourne & (X) & 69.17 & False
\end{tabularx}
\caption{Results of the trials using the Anderson data set. For the reduced data, we picked points within $\delta =2$ of the hyperplane learned using a linear SVM. Included are two separate runs on the reduced, randomized data with a real 16-qubit computer. The second run did not have time measured.}
\end{table*}

\begin{table}[!htbp]
\tiny
\begin{tabularx}{\textwidth}{c|l|l|X|X|X}
\textbf{Data} & \textbf{Algorithm} & \textbf{Backend} & \textbf{Time (Seconds, rounded)} & \textbf{Accuracy (\%)} & \textbf{Only one Class Predicted?} \\ \hline
\multirow{5}{*}{All Data} & SVM, Linear Kernel, scaled & Local Processor & 0 & 90.56 & False \\
 & SVM, RBF Kernel, scaled & Local Processor & 0 & 90.99 & False \\
 & QSVM & qasm\_simulator & 28505 & 81.02 & True \\
 & VQC & qasm\_simulator & 2825 & 61.29 & False \\ 
 & QSVM & ibmq\_16\_melbourne & (X) & 81.02 & True \\ \hline
\multirow{4}{*}{\begin{tabular}[c]{@{}c@{}}All Data, \\ Randomized\end{tabular}} & SVM, Linear Kernel, scaled & Local Processor & 0 & 81.02 & True \\
 & SVM, RBF Kernel, scaled & Local Processor & 0 & 90.35 & False \\
 & QSVM & qasm\_simulator & 28562 & 81.02 & True \\
 & VQC & qasm\_simulator & 2845 & 65.64 & False \\ \hline
\multirow{4}{*}{Reduced Data} & SVM, Linear Kernel, scaled & Local Processor & 0 & 92.14 & False \\
 & SVM, RBF Kernel, scaled & Local Processor & 0 & 91.92 & False \\
 & QSVM & qasm\_simulator & 27485 & 84.61 & True \\
 & VQC & qasm\_simulator & 2709 & 61.13 & False \\ \hline
\multirow{4}{*}{\begin{tabular}[c]{@{}c@{}}Reduced Data, \\ Randomized\end{tabular}} & SVM, Linear Kernel, scaled & Local Processor & 0 & 84.61 & True \\
 & SVM, RBF Kernel, scaled & Local Processor & 0 & 90.92 & False \\
 & QSVM & qasm\_simulator & 28533 & 54.04 & False \\
 & VQC & qasm\_simulator & 2826 & 64.78 & False
\end{tabularx}
\caption{Results of Coronavirus trials where the cutoff for safety is 5 new cases per day. For the reduced data, we picked points within $\delta =2000$ of the hyperplane learned using a linear SVM. Time was not measured for the 16-qubit run.}
\end{table}

\begin{table*}[!htbp]
\tiny
\begin{tabularx}{\textwidth}{c|l|l|X|X|X}
\textbf{Data} & \textbf{Algorithm} & \textbf{Backend} & \textbf{Time (Seconds, rounded)} & \textbf{Accuracy (\%)} & \textbf{Only one Class Predicted?} \\ \hline
\multirow{4}{*}{All Data} & SVM, Linear Kernel, scaled & Local Processor & 0 & 87.92 & False \\
 & SVM, RBF Kernel, scaled & Local Processor & 0 & 87.92 & False \\
 & QSVM & qasm\_simulator & 27450 & 74.89 & True \\
 & VQC & qasm\_simulator & 3003 & 58.9 & False \\ \hline
\multirow{4}{*}{\begin{tabular}[c]{@{}c@{}}All Data, \\ Randomized\end{tabular}} & SVM, Linear Kernel, scaled & Local Processor & 0 & 74.89 & True \\
 & SVM, RBF Kernel, scaled & Local Processor & 0 & 86.33 & False \\
 & QSVM & qasm\_simulator & 28466 & 74.89 & True \\
 & VQC & qasm\_simulator & 3000 & 61.12 & False \\ \hline
\multirow{4}{*}{Reduced Data} & SVM, Linear Kernel, scaled & Local Processor & 0 & 89.54 & False \\
 & SVM, RBF Kernel, scaled & Local Processor & 0 & 90.66 & False \\
 & QSVM & qasm\_simulator & 26579 & 78.53 & True \\
 & VQC & qasm\_simulator & 2869 & 59.96 & False \\ \hline
\multirow{4}{*}{\begin{tabular}[c]{@{}c@{}}Reduced Data, \\ Randomized\end{tabular}} & SVM, Linear Kernel, scaled & Local Processor & 0 & 78.53 & True \\
 & SVM, RBF Kernel, scaled & Local Processor & 0 & 86.21 & False \\
 & QSVM & qasm\_simulator & 27890 & 78.53 & True \\
 & VQC & qasm\_simulator & 2849 & 64.07 & False
\end{tabularx}
\caption{Results of Coronavirus trials where the cutoff for safety is 3 new cases per day. For the reduced data, we picked points within $\delta =2000$ of the hyperplane learned using a linear SVM.}
\end{table*}

\begin{table*}[!htpb]
\tiny
\begin{tabularx}{\textwidth}{c|l|l|X|X|X}
\textbf{Data} & \textbf{Algorithm} & \textbf{Backend} & \textbf{Time (Seconds, rounded)} & \textbf{Accuracy (\%)} & \textbf{Only one Class Predicted?} \\ \hline
\multirow{4}{*}{All Data} & SVM, Linear Kernel, scaled & Local Processor & 0 & 80.51 & False \\
 & SVM, RBF Kernel, scaled & Local Processor & 0 & 79.77 & False \\
 & QSVM & qasm\_simulator & 29125 & 60.06 & True \\
 & VQC & qasm\_simulator & 3005 & 52.86 & False \\ \hline
\multirow{4}{*}{\begin{tabular}[c]{@{}c@{}}All Data, \\ Randomized\end{tabular}} & SVM, Linear Kernel, scaled & Local Processor & 0 & 60.06 & True \\
 & SVM, RBF Kernel, scaled & Local Processor & 0 & 78.5 & False \\
 & QSVM & qasm\_simulator & 30751 & 60.06 & True \\
 & VQC & qasm\_simulator & 3014 & 54.77 & False \\ \hline
\multirow{4}{*}{Reduced Data} & SVM, Linear Kernel, scaled & Local Processor & 0 & 83.2 & False \\
 & SVM, RBF Kernel, scaled & Local Processor & 0 & 84.66 & False \\
 & QSVM & qasm\_simulator & 26656 & 63.49 & True \\
 & VQC & qasm\_simulator & 2855 & 52.86 & False \\ \hline
\multirow{4}{*}{\begin{tabular}[c]{@{}c@{}}Reduced Data, \\ Randomized\end{tabular}} & SVM, Linear Kernel, scaled & Local Processor & 0 & 63.49 & True \\
 & SVM, RBF Kernel, scaled & Local Processor & 0 & 76.48 & False \\
 & QSVM & qasm\_simulator & 29351 & 63.49 & True \\
 & VQC & qasm\_simulator & 2775 & 57.11 & False
\end{tabularx}
\caption{Results of Coronavirus trials where the cutoff for safety is 1 new case per day. For the reduced data, we picked points within $\delta =2000$ of the hyperplane learned using a linear SVM.}
\end{table*}












\end{document}